\documentclass[letterpaper,12pt]{article}
\usepackage{authblk}
\usepackage[utf8]{inputenc}
\usepackage[english]{babel}
\usepackage{amsmath}
\usepackage{amssymb}
\usepackage{graphicx}
\usepackage{multirow}
\usepackage{epsfig}
\usepackage{rotating}
\usepackage{longtable} 
\usepackage{caption}
\usepackage{subcaption}
\usepackage{url}
\usepackage{pdflscape}
\usepackage{lscape}
\usepackage{cancel}
\usepackage{setspace}
\pdfoutput=1
\usepackage{booktabs}
\usepackage{csquotes}
\usepackage{placeins}
\usepackage[left=2cm,top=3cm,right=2cm,bottom=3cm]{geometry}
\usepackage{tabularx}
\usepackage{adjustbox}
\usepackage{color,xcolor}

\usepackage{hyperref}

\usepackage{mathtools}
\usepackage{latexsym}

\title{One texture zero for Dirac neutrinos in a diagonal charged lepton basis}

\author[a]{Richard H. Benavides}
\author[a]{John D. Gómez}
\author[b]{William A. Ponce}

\affil[a]{Departamento de Educación y Ciencias Básicas, Facultad de Ciencias Exactas y Aplicadas, Institución Universitaria ITM, Medell\'{i}n, Colombia.} 
\affil[b]{Instituto de Física, Universidad de Antioquia, A.A. 1226, Medellín, Colombia.}
\vspace{1cm}

\affil[a]{{Corresponding author:richardbenavides@itm.edu.co}}

\begin{document}

\title{Three Texture Zeros in the Hermitian Dirac sector within Type-I Seesaw Mechanism}

\maketitle

\begin{abstract}

We present numerical benchmark solutions for Hermitian Dirac neutrino mass matrices with three texture zeros within the type-I seesaw mechanism. The analysis is  to normal ordering (NO) and inverted (IO). The heavy Majorana mass matrix is assumed diagonal, while the Dirac mass matrix is taken Hermitian. Four viable textures in NO are shown to reproduce the neutrino oscillation observables within the $3\sigma$ ranges of the NuFIT global analysis, and three viable textures in IO. 
\end{abstract}

\section{Introduction}

The Standard Model (SM) of particle physics remains the most successful theoretical framework for describing the fundamental interactions of elementary particles \cite{Glashow:1961tr,Weinberg:1967tq,Salam:1968rm}. Despite its remarkable predictive power and experimental success, the SM is unable to explain several well-established phenomena, among them the existence of dark matter and the nonzero masses and oscillations of neutrinos \cite{Bertone:2004pz,Super-Kamiokande:1998kpq,SNO:2002tuh}.

Neutrino oscillation experiments have conclusively demonstrated that neutrinos are massive particles and that leptonic flavors mix during propagation \cite{Kajita:2016cak,McDonald:2016ixn,Esteban:2020cvm}. Consequently, the generation of neutrino masses constitutes one of the most important indications of physics beyond the Standard Model. Among the various mechanisms proposed to explain the smallness of neutrino masses, the Type-I seesaw mechanism is one of the simplest and most elegant scenarios \cite{Minkowski:1977sc,Yanagida:1980xy,Gell-Mann:1979vob,Mohapatra:1979ia,Schechter:1980gr}. In this framework, the light neutrino masses emerge naturally through the introduction of heavy right-handed Majorana neutrinos.

Motivated by current neutrino oscillation data, this work investigates texture-zero structures in both normal ordering (NO) and inverted ordering (IO) schemes. Texture zeros provide a useful phenomenological approach for reducing the number of free parameters in fermion mass matrices while preserving predictive power \cite{Frampton:2002yf,Xing:2002ta,Fritzsch:1999ee}. In particular, we consider a general $6\times6$ neutrino mass matrix consistent with the Type-I seesaw mechanism, composed of a vanishing $3\times3$ block in the upper-left sector, a Dirac-type submatrix together with its transpose in the off-diagonal sectors, and a Majorana-type submatrix in the lower-right diagonal block. The complete mass matrix can therefore be written as

\begin{equation}
M_\nu=
\begin{pmatrix}
0 & M_D \\
M_D^T & M_R
\end{pmatrix},
\label{masa}
\end{equation}

\noindent where $M_D$ corresponds to the Dirac neutrino mass matrix and $M_R$ denotes the heavy Majorana mass matrix.

Our main hypothesis consists in assuming that the Dirac mass matrix is Hermitian. This condition guarantees that the off-diagonal texture zeros appear symmetrically and are counted only once, significantly reducing the number of independent free parameters \cite{PhysRevD.76.013002,Fritzsch:1999ee}. Under this assumption, it becomes possible to construct viable mass textures containing at most three texture zeros compatible with current experimental observables.

As shown in this work, among the twenty possible three-zero texture configurations, only four are compatible with experimental data in the normal ordering case, while three viable solutions are found for inverted ordering. Additionally, throughout this analysis the Majorana mass matrix is assumed to be diagonal as usual. The resulting textures successfully reproduce the observed neutrino mixing angles, mass-squared differences, and other relevant phenomenological constraints.

\section{The model}

We begin from the possibility to construct a neutrino mass matrix capable of generating the Type-I seesaw mechanism. In particular, we assume that the Dirac neutrino mass matrix can be taken Hermitian, while the Majorana mass matrix is considered diagonal in the flavor basis. Additionally, the upper-left $3\times3$ block of the complete neutrino mass matrix is assumed to contain vanishing entries \cite{Minkowski:1977sc,Mohapatra:1979ia,Schechter:1980gr}. Under these assumptions, the complete neutrino mass matrix takes the form \ref{masa}, where $M_D$ denotes the Dirac mass matrix and $M_R$ corresponds to the Majorana mass matrix of the heavy right-handed neutrinos.

Since the present analysis is based exclusively on the phenomenological structure of the mass matrices, our results are independent of the underlying gauge model. Therefore, the study presented here can be interpreted as a model-independent analysis of texture-zero structures within the Type-I seesaw framework \cite{Frampton:2002yf,Xing:2002ta,Fritzsch:1999ee}.

The Hermitian condition imposed on the Dirac mass matrix considerably reduces the number of independent parameters and guarantees that the texture zeros outside the diagonal appear symmetrically. This assumption enhances the predictive power of the framework and allows a systematic exploration of viable texture-zero configurations compatible with current neutrino oscillation data.

In the type-I seesaw mechanism, the effective light neutrino mass matrix is given by 
\begin{equation}\label{SeeSaw}
m_\nu = - M_D M_R^{-1} M_D^T,
\end{equation}
where $M_D$ is the Dirac mass matrix, $M_R$ is the Majorana mass matrix, and the minus sign is a direct result of the linear algebra used to diagonalize the overall \(6 \times 6\) neutrino mass matrix; the absolute overall sign of a mass matrix is not physically observable on its own because it can be absorbed.

Our working frame is the SM enlarged with three right-handed neutrinos, one for each lepton flavor. We also assume a charged lepton mass matrix diagonal in order to properly use the phenomenological experimental values.   Also, to simplify matters and reduce the number of free parameters, we assume a Hermitian Dirac mass matrix $M_D$ and a diagonal Majorana mass matrix $M_R$.

A Hermitian $3\times3$ matrix has six independent entries:
\[
(11), (22), (33), (12), (13), (23).
\]

As the general expression for the number of possibilities to obtain texture matrices with three texture zeros is given by the binomial coefficient, 
\begin{equation}
\binom{n}{k}
=
\frac{n!}{k!(n-k)!},
\end{equation}
where \(n\) represents the total number of available elements and \(k\) denotes the number of selected elements.

In the present case, since the Hermitian Dirac mass matrix contains six independent entries and we choose three of them to be texture zeros, the total number of possible configurations is
\begin{equation}
\binom{6}{3}
=
\frac{6!}{3!(6-3)!}
=
20,
\end{equation}

\noindent distinct textures. In this work, we analyze these twenty possible texture configurations containing three texture zeros in the Dirac neutrino mass matrix and study their compatibility with current experimental neutrino oscillation data \cite{Esteban:2020cvm,NuFIT}. The analysis is performed for both normal and inverted mass ordering schemes, taking into account the observed neutrino mixing angles, mass-squared differences, and phenomenological constraints on the neutrino sector.

\section{Observables}

The physical observables considered in this analysis are the three leptonic mixing angles, the neutrino mass-squared differences, and the Dirac CP-violating phase, as reported by the NuFIT global analysis \cite{NuFIT,Esteban:2020cvm}. For normal and inverted ordering (NO, IO), we consider these observables to constitute the experimental constraints used to determine the viability of each texture-zero configuration within the Type-I seesaw framework.

Mixing angles:
\begin{align}
\sin^2\theta_{13} = |U_{e3}|^2, \,\,\,\,
\sin^2\theta_{12} = \frac{|U_{e2}|^2}{1-|U_{e3}|^2}, \,\,\,\,
\sin^2\theta_{23} = \frac{|U_{\mu3}|^2}{1-|U_{e3}|^2},
\end{align}
where we have used the unitarity property of the $U_{PMNS} $ matrix.\\

Mass differences:
\begin{align}
\Delta m_{21}^2 = m_2^2 - m_1^2, \,\,\,\,
\Delta m_{3\ell}^2 =
\begin{cases}
m_3^2 - m_1^2 & (\text{NO})\\
m_3^2 - m_2^2 & (\text{IO})
\end{cases}.
\end{align}

Jarlskog invariant:
\begin{equation}
J_{CP} = \text{Im}(U_{e1}U_{\mu2}U_{e2}^*U_{\mu1}^*).
\label{jar}
\end{equation}

These relations are required for our analysis.

\section{Neutrinoless Double Beta Decay ($0\nu\beta\beta$)}

The search for neutrinoless double beta decay ($0\nu\beta\beta$) constitutes the most sensitive experimental probe to determine the Majorana nature of neutrinos and to constrain the absolute neutrino mass scale, serving as a powerful test for physics beyond the SM \cite{Furry:1939qr, Schechter:1981bd, Bilenky:2014uka, Dolinski:2019zjk}. In the standard flavor basis, defined as the representation where the charged-lepton mass matrix $M_{\ell}$ is diagonal, the decay amplitude is directly proportional to the effective Majorana mass, denoted as $m_{\beta\beta}$. Under this specific basis choice, $m_{\beta\beta}$ corresponds identically to the magnitude of the $(1,1)$ element of the light neutrino mass matrix in the flavor basis, $m_{\nu}$: 

\[m_{\beta\beta} \equiv \left| (m_{\nu})_{11} \right| = \left| \sum_{i=1}^{3} U_{1i}^2 m_i \right|,\]

\noindent where $m_i$ are the absolute masses of the light neutrino autostates ($i=1,2,3$), and $U_{1i}$ represent the elements of the first row of the PMNS lepton mixing matrix \cite{Pontecorvo:1957cp, Maki:1962mu}. Explicitly, expanding in terms of the mixing angles ($\theta_{12}, \theta_{13}$) and the Dirac CP-violating phase ($\delta_{\text{CP}}$), the effective mass is given by:

\[m_{\beta\beta} = \left| m_1 c_{12}^2 c_{13}^2 + m_2 s_{12}^2 c_{13}^2 e^{i\alpha} + m_3 s_{13}^2 e^{i(\beta - 2\delta_{\text{CP}})} \right|,\] 

\noindent here, $c_{ij} = \cos\theta_{ij}$ and $s_{ij} = \sin\theta_{ij}$, while $\alpha$ and $\beta$ stand for the Majorana CP-violating phases.In our model, the light neutrino mass matrix $m_{\nu}$ is generated via a Type-I See-saw mechanism governed by the relation  \ref{SeeSaw}. Since we work under the standard phenomenological assumption that the charged-lepton sector is already diagonalized, the full structural output of the see-saw mechanism maps directly onto the flavor basis. Consequently, by imposing a Hermitian Dirac mass matrix $M_D$ with three textures zeros, the effective Majorana mass is completely determined by the absolute value $(1,1)$ entry of the calculated matrix, meaning that $m_{\beta\beta} \equiv |(m_{\nu})_{11}|$ is entirely dictated by the high-scale parameters ($a, b, c, d, f$) and the heavy Majorana matrix $M_R$.

\section{Classification of Textures}

Following the methodology developed in previous works on neutrino texture zeros \cite{Benavides:2020pjx,Lenis:2023lgq}, we determine the corresponding neutrino mass matrices and obtain the unitary matrices that properly diagonalize them. Subsequently, a statistical $\chi^2$ analysis is performed in order to fit the three leptonic mixing angles, the two neutrino mass-squared differences, and the Jarlskog invariant, from which the Dirac CP-violating phase is extracted.

All observables are required to lie within the $3\sigma$ ranges reported by the NuFIT global analysis \cite{NuFIT}. In this way, each texture-zero configuration is tested against current experimental constraints on neutrino oscillation parameters.

After analyzing the twenty possible three-zero texture configurations in both normal and inverted ordering scenarios, only the following seven textures were found to be phenomenologically viable, three of which are compatible with both scenarios:

\begin{center}
\begin{tabular}{ |c|c| } 
 \hline
 NO & IO\\
 \hline
 T1: $(11),(22),(33)$ & T1: $(11),(22),(33)$\\
 T2: $(11),(22),(23)$ &T2: $(11),(22),(23)$\\
 T3: $(11),(33),(12)$ & T4: $(11),(33),(23)$\\
 T4: $(11),(33),(23)$ & \\
 \hline
\end{tabular}
\end{center}
\subsection{$T_1$ Normal Ordering }

In this particular case, the three texture zeros are located along the diagonal of the Dirac neutrino mass matrix, which can be written in the form
\begin{equation}
M_D=
\begin{pmatrix}
0 & a+i b & c+id \\
a-ib & 0 & e+if \\
c-id & e-if & 0
\end{pmatrix},\label{MdT1}
\end{equation}
where, due to the Hermitian condition imposed on the Dirac sector,
the number of independent free parameters is significantly reduced, increasing the predictive capability of the framework.

The effective light-neutrino mass matrix is obtained through the diagonalization relation
\begin{equation}
U_{\text{PMNS}}^{T}\, m_{\nu}\, U_{\text{PMNS}}
=
\mathrm{diag}(m_1,m_2,m_3),
\end{equation}
where \(U_{\text{PMNS}}\) denotes the Pontecorvo--Maki--Nakagawa--Sakata (PMNS) leptonic mixing matrix \cite{Pontecorvo:1957cp,Maki:1962mu}. The parameters \(m_1\), \(m_2\), and \(m_3\) correspond to the physical masses of the light neutrino states. The previous relation is valid because throughout this work we consider the basis in which the charged-lepton mass matrix is diagonal. Consequently, all leptonic flavor mixing is entirely encoded in the PMNS matrix, which directly diagonalizes the effective light-neutrino mass matrix.

\subsection{Numerical Simulation and Leptonic CP-Violating Phase Analysis}
\subsubsection{Numerical Method and Parameter Scanning}
To evaluate the phenomenological viability of the proposed Hermitian Dirac mass matrix with a vanishing diagonal (null-diagonal texture) \cite{Frampton:2002yf, Xing:2002ta}, we performed a comprehensive numerical scan using an optimized Monte Carlo routine in Mathematica. The structural ansatz for the Dirac sector is parameterized as equation \ref{MdT1},where $a, b, c, d, e, f$ are six free real parameters sampled randomly over wide intervals centered around their optimal global minima. The heavy right-handed Majorana neutrino mass matrix is fixed at the typical high-energy see-saw scale\cite{Minkowski:1977sc}, $M_R = \text{diag}(1.000, 5.042, 5.047) \times 10^{14}\text{ GeV}$ found in our analysis.
For each generated parameter set, the full light neutrino mass matrix is numerically evaluated via the Type-I See-saw relation, $m_{\nu} = -M_D M_R^{-1} M_D^T$. Throughout this work, we operate under the standard phenomenological framework where the charged-lepton mass matrix $M_{\ell}$ is already diagonalized. Consequently, the computed $m_{\nu}$ directly represents the light neutrino mass matrix in the flavor basis.

By computing the eigenvalues and eigenvectors of the hermitian combination $h = m_{\nu}m_{\nu}^{\dagger}$, we extract the theoretical neutrino mass squared differences ($\Delta m_{21}^2$, and $\Delta m_{31}^2$ or $\Delta m_{32}^2$), the absolute mass of the lightest neutrino state ($m_1$ for Normal Hierarchy and $m_3$ for Inverted Hierarchy), and the PMNS mixing parameters ($s_{12}^2, s_{13}^2, s_{23}^2$), as well as the CP-violating phase. The latter is obtained through the Jarlskog invariant, defined as equation \ref{jar} \cite{Jarlskog:1985ht}. Each simulated configuration is statistically filtered using a global-fit $\chi^2$ penalization function based on the latest NuFIT 6.0 data releases \cite{Esteban:2020cvm}:

\[\chi^2 = \sum_{k} \frac{(\mathcal{O}_k^{\text{theo}} - \mathcal{O}_k^{\text{exp}})^2}{\sigma_k^2},\]

\noindent where $\mathcal{O}_k$ and $\sigma_k$ represent the neutrino oscillation observables and their respective experimental $1\sigma$ uncertainties \cite{Esteban:2020cvm}.
\noindent
To determine the confidence intervals for a joint estimation of two independent oscillation parameters (e.g., mixing angles or mass squared differences), we evaluate the profile likelihood using the delta chi-squared statistic:
\begin{equation}
    \Delta\chi^2(\mathbf{p}) = \chi^2(\mathbf{p}) - \chi^2_{\text{min}},
\end{equation}
where $\mathbf{p}$ represents the vector of the two parameters of interest, and $\chi^2_{\text{min}}$ is the global minimum obtained by varying all parameters.

Under the assumption of asymptotic normality (Wilks' theorem), $\Delta\chi^2$ follows a chi-squared distribution with $k = 2$ degrees of freedom ($k=2$ d.o.f.). The cumulative probability $P$ for a given threshold $\Delta\chi^2$ is defined by:
\begin{equation}
    P(\Delta\chi^2 \le \Delta\chi^2_0) = 1 - e^{-\Delta\chi^2_0 / 2}.
\end{equation}

Consequently, the boundaries for the standard joint confidence regions are defined by the following specific critical values:
\begin{itemize}
    \item \textbf{$1\sigma$ region} ($68.27\%$ C.L.): $\Delta\chi^2 < 2.30$
    \item \textbf{$2\sigma$ region} ($95.45\%$ C.L.): $\Delta\chi^2 < 6.18$
    \item \textbf{$3\sigma$ region} ($99.73\%$ C.L.): $\Delta\chi^2 < 11.83$
\end{itemize}
These thresholds define the multi-dimensional confidence ellipses when projecting the multi-parameter space onto any two-dimensional parameter plane.
Therefore, accepted points are subsequently classified into strict confidence levels: the $1\sigma$ region ($\Delta\chi^2 < 2.30$), the $2\sigma$ region ($\Delta\chi^2 < 6.18$), and the $3\sigma$ region ($\Delta\chi^2 < 11.83$).




In this way, the best-fit regions are determined by minimizing the $\chi^2$ function and comparing the analytical predictions of the texture-zero framework with the current experimental neutrino oscillation data.

The numerical results obtained for this first texture are the following:

\begin{equation}
M_R =
\begin{pmatrix}
1.000\times10^{14} & 0 & 0 \\
0 & 5.042\times10^{14} & 0 \\
0 & 0 & 5.047\times10^{14}
\end{pmatrix} \text{ GeV},
\end{equation}

\begin{equation}
M_D =
\begin{pmatrix}
0 & 17.5781+51.9210i & -36.8446+15.2568i \\
17.5781-51.9210i & 0 & 56.9667-25.3268i \\
-36.8446-15.2568i & 56.9667+25.3268i & 0
\end{pmatrix} \text{ GeV}.
\end{equation}

\begin{align}
\sin^2\theta_{12} &\approx 0.304, \,\,\,\,  \sin^2\theta_{13} \approx 0.022, \,\,\,\, 
\sin^2\theta_{23} \approx 0.570, \\
\Delta m_{21}^2 &\approx 7.42\times10^{-5}\ \text{eV}^2, \,\,\,
\Delta m_{31}^2 \approx 2.51\times10^{-3}\ \text{eV}^2,\,\,\,
\delta_{CP} \approx 202^\circ. \nonumber
\end{align}

With this numerical analysis, the resulting light-neutrino masses are obtained as
\begin{equation}
(m_1,m_2,m_3)\approx(0.000801,\ 0.008651,\ 0.050106)\ \text{eV},
\end{equation}
which are also consistent with the current upper bounds on the sum of the three active neutrino masses imposed by cosmological observations \cite{Planck:2018vyg,Vagnozzi:2017ovm}. These cosmological constraints provide an important complementary test for the viability of neutrino mass models and texture-zero scenarios. This cosmological upper bound on the sum of the three active neutrino masses is,
\begin{equation}
\sum_i m_{\nu_i} < 0.12~\text{eV},
\end{equation}
obtained from cosmological observations within the $\Lambda$CDM framework \cite{Planck:2018vyg,DiValentino:2021hoh}.

\subsubsection{CP-Violating Phase Projections}
Leptonic CP violation is inherently quantified within the PMNS framework by the Jarlskog invariant, $J_{\text{CP}}$, which is determined from the mixing matrix elements as equation       \ref{jar}. For every statistically viable point generated by the see-saw mechanism, the Dirac CP-violating phase $\delta_{\text{CP}}$ is extracted numerically using the reconstructed mixing angles and the value of $J_{\text{CP}}$. To present a complete and transparent phenomenological mapping, the generated data points are simultaneously projected onto two distinct diagnostic planes:
\begin{figure}[h]
\begin{subfigure}{0.5\textwidth}
\includegraphics[width=0.9\linewidth, height=6cm]{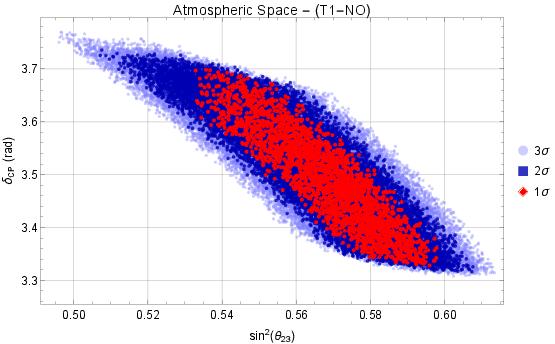} 
\caption{ The $(\theta_{23} \text{ vs. } \delta_{\text{CP}})$ Plane}
\label{fig:subim1}
\end{subfigure}
\begin{subfigure}{0.5\textwidth}
\includegraphics[width=0.9\linewidth, height=6cm]{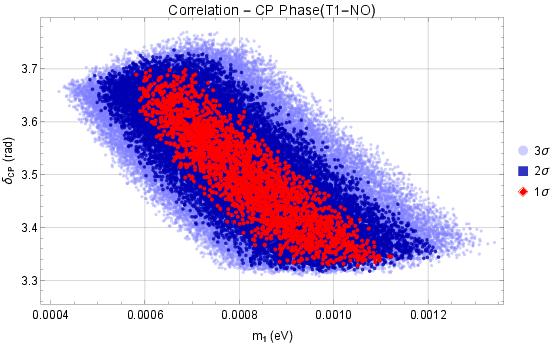}
\caption{The $(m_{\text{lightest}} \text{ vs. } \delta_{\text{CP}})$ Plane}
\label{fig:subim2}
\end{subfigure}
\caption{In the left figure, we have the $\theta_{23}$  mixing angle vs. CP phase, and in the right figure, we have the \(m_{\text{lightest}}\) vs. the CP phase.}
\label{fig:image2}
\end{figure}

\begin{itemize}
    \item The $(\theta_{23} \text{ vs. } \delta_{\text{CP}})$ Plane: This projection is crucial to assess the correlation between the CP-violating phase and the octant behavior of the atmospheric mixing angle (i.e., whether $\theta_{23} < 45^\circ$ or $\theta_{23} > 45^\circ$). It explicitly showcases how the constrained structure of the null-diagonal texture bounds the permissible values of $\delta_{\text{CP}}$ across different confidence intervals.
    \item The $(m_{\text{lightest}} \text{ vs. } \delta_{\text{CP}})$ Plane: This mapping displays the evolution of the CP-violating phase as a function of the absolute mass scale ($m_1$ for Normal Hierarchy and $m_3$ for Inverted Hierarchy). It reveals the predictive narrow bands or isolated islands allowed by the model, illustrating how tightly the texture controls the phase in the limit of a vanishing lightest neutrino mass.
\end{itemize}

\subsection{Phenomenological Predictions in the Neutrinoless Double Beta Decay}
\begin{figure}[h]
\centering
 \includegraphics[width=11cm]{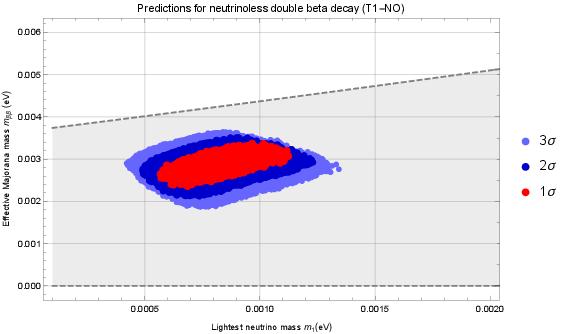}
 \caption{Allowed regions for the effective Majorana neutrino mass $m_{\beta\beta}$ as a function of the lightest neutrino mass.}
 \label{fig:mbbT1}
 \end{figure}

 The filtered data points originating from our Monte Carlo simulation are projected onto the $(m_{\text{lightest}}, m_{\beta\beta})$ plane and superimposed onto the standard analytical phenomenological bands (Fig. \ref{fig:mbbT1}). The background gray shaded area represents the unconstrained, model-independent phenomenological region standardly allowed for the Normal Ordering (NO) scenario. This target band is analytically derived by varying the standard neutrino oscillation parameters ($\theta_{12}$, $\theta_{13}$, and $\Delta m_{21}^2, \Delta m_{31}^2$) within their globally accepted $3\sigma$ experimental confidence intervals, while allowing the unmeasured Majorana CP-violating phases ($\alpha$ and $\beta$) to scan freely across their full theoretical range $[0, 2\pi]$.

By superimposing our simulation results, it becomes visually manifest how the specific T1 Dirac texture under study introduces tight, localized algebraic constraints. Rather than populating the entire gray landscape, the model restricts the allowed parameters to a highly predictive, compact island well within the standard boundaries, significantly reducing the phenomenological uncertainty of $m_{\beta\beta}$ for a precise range of the lightest neutrino mass $m_1$. This allows us to contrast the predictive power of our specific see-saw texture directly against current upper limits established by the KamLAND-Zen collaboration \cite{KamLAND-Zen:2022gqc} ($m_{\beta\beta} < 36-156 \text{ meV}$) and the future discovery sensitivities of upcoming next-generation $0\nu\beta\beta$ experimental setups, such as LEGEND \cite{LEGEND:2021bnm}. In the case of the Normal Ordering (NO), the model successfully maps the conditions under which the effective Majorana mass drops toward near-zero values due to destructive quantum interference, while for the Inverted Hierarchy (IO), it outlines how the parameters populate the strict horizontal physical lower bound ($m_{\beta\beta} \gtrsim 15\text{ meV}$), providing clear experimental benchmarks for model exclusion.

 \subsection{Additional Viable Textures in NO}

Proceeding in the same way as in the previous analysis for the normal ordering configurations, we find three additional texture-zero structures that successfully reproduce the current experimental neutrino oscillation data. These viable configurations are summarized in the following table \ref{Table1}, while a more detailed numerical analysis and the corresponding best-fit parameters are presented in the Appendix A.

\begin{table}[h]
\centering
\caption{Summary of viable normal-ordering benchmark points.}
\begin{tabular}{c c c c c c c c}
\toprule
Texture & $\sin^2\theta_{12}$ & $\sin^2\theta_{13}$ & $\sin^2\theta_{23}$ & $\delta_{CP}$ & $\Delta m_{21}^2$ & $\Delta m_{31}^2$ & $\chi_{min}^2$\\
\midrule
T1 & 0.304 & 0.0222 & 0.5700 & $202.03^\circ$ & $7.42\times10^{-5}$ & $2.51\times10^{-3}$ & $1.25\times 10^{-14}$\\
T2 & 0.303 & 0.0219 & 0.5374 & $264.8
8^\circ$ & $7.45\times10^{-5}$ & $2.51\times10^{-3}$ & $1.53$ \\
T3 & 0.304 & 0.0222 & 0.5700 & $231.42^\circ$ & $7.42\times10^{-5}$ & $2.51\times10^{-3}$ & $9.37\times10^{-10}$ \\
T4 & 0.306 & 0.0221 & 0.5692 & $182.00^\circ$ & $7.49\times10^{-5}$ & $2.51\times10^{-3}$ & $0.22$\\
\bottomrule
\end{tabular}
\label{Table1}
\end{table}

Although Textures T1 and T3, absolute minima exhibit an excellent agreement with experimental data ($\chi^2 \approx 0$), their allowed parameter spaces are highly constrained, yielding a lower efficiency in the random Monte Carlo sampling ($\sim 0.9\%$ acceptance rate over $10^7$ points). This indicates a higher degree of fine-tuning required among the high-scale Dirac parameters. Conversely, Texture T2 and T4, exhibits a residual minimum ($\chi^2 \approx 1.5$ and $\chi^2 \approx 0.22$) but provides a significantly larger volume of phenomenologically viable solutions ($\sim 3\%$ and $\sim 1.4\%$ acceptance rate, respectively), highlighting a robust and less fine-tuned parameter landscape under current global oscillation fits. The results are shown in Figure 3.

\begin{figure*}[p] 
    \centering
    
    \begin{subfigure}[b]{0.32\textwidth}
        \centering
        \includegraphics[width=\textwidth]{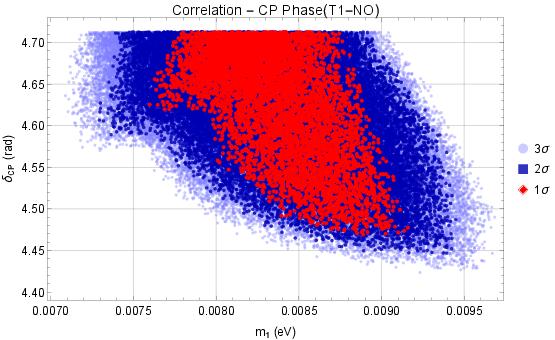}
        \caption{T2: $\delta_{\text{CP}}$ vs. $m_1$.}
        \label{fig:T2_cp_m1}
    \end{subfigure}
    \hfill
    \begin{subfigure}[b]{0.32\textwidth}
        \centering
        \includegraphics[width=\textwidth]{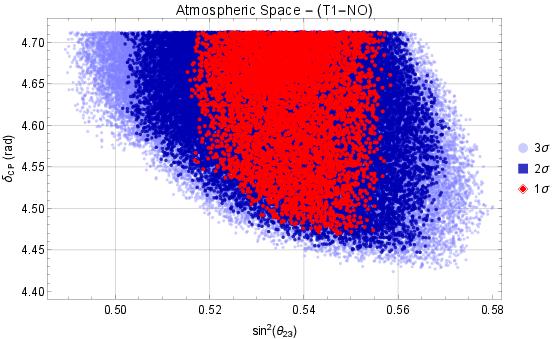}
        \caption{T2: $\sin^2\theta_{23}$ vs. $\delta_{\text{CP}}$.}
        \label{fig:T2_cp_t23}
    \end{subfigure}
    \hfill
    \begin{subfigure}[b]{0.32\textwidth}
        \centering
        \includegraphics[width=\textwidth]{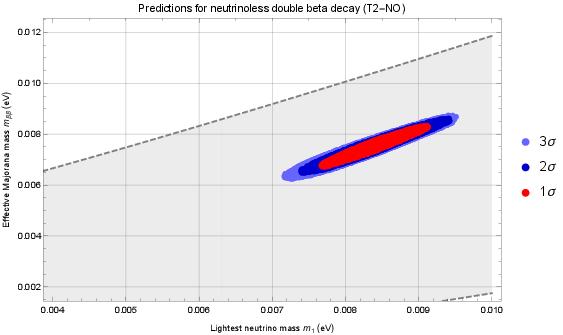}
        \caption{T2: $m_{\beta\beta}$ predictions.}
        \label{fig:T2_db}
    \end{subfigure}
    
    \vspace{0.3cm} 
    
    \begin{subfigure}[b]{0.32\textwidth}
        \centering
        \includegraphics[width=\textwidth]{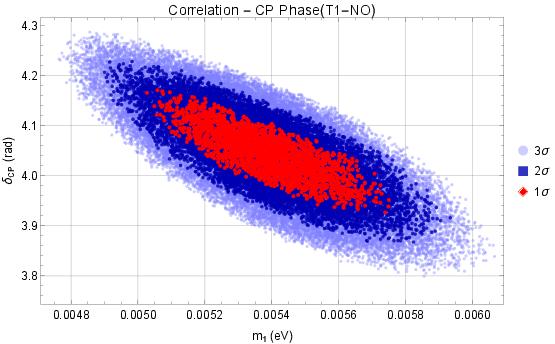}
        \caption{T3: $\delta_{\text{CP}}$ vs. $m_1$.}
        \label{fig:T3_cp_m1}
    \end{subfigure}
    \hfill
    \begin{subfigure}[b]{0.32\textwidth}
        \centering
        \includegraphics[width=\textwidth]{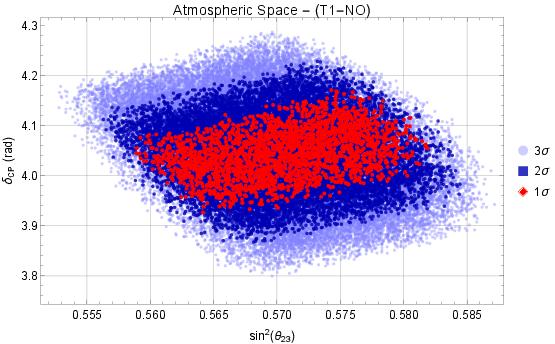}
        \caption{T3: $\sin^2\theta_{23}$ vs. $\delta_{\text{CP}}$.}
        \label{fig:T3_cp_t23}
    \end{subfigure}
    \hfill
    \begin{subfigure}[b]{0.32\textwidth}
        \centering
        \includegraphics[width=\textwidth]{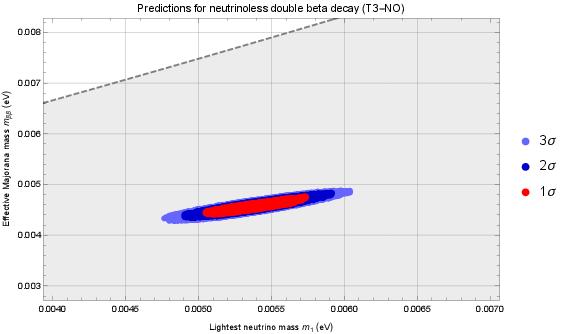}
        \caption{T3: $m_{\beta\beta}$ predictions.}
        \label{fig:T3_db}
    \end{subfigure}
    
    \vspace{0.3cm} 
    
    \begin{subfigure}[b]{0.32\textwidth}
        \centering
        \includegraphics[width=\textwidth]{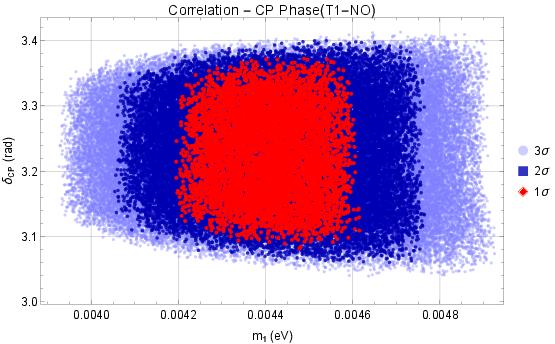}
        \caption{T4: $\delta_{\text{CP}}$ vs. $m_1$.}
        \label{fig:T4_cp_m1}
    \end{subfigure}
    \hfill
    \begin{subfigure}[b]{0.32\textwidth}
        \centering
        \includegraphics[width=\textwidth]{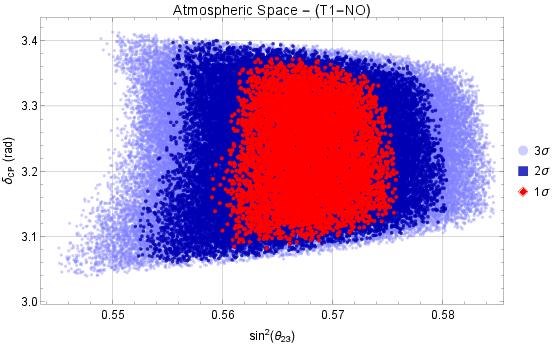}
        \caption{T4: $\sin^2\theta_{23}$ vs. $\delta_{\text{CP}}$.}
        \label{fig:T4_cp_t23}
    \end{subfigure}
    \hfill
    \begin{subfigure}[b]{0.32\textwidth}
        \centering
        \includegraphics[width=\textwidth]{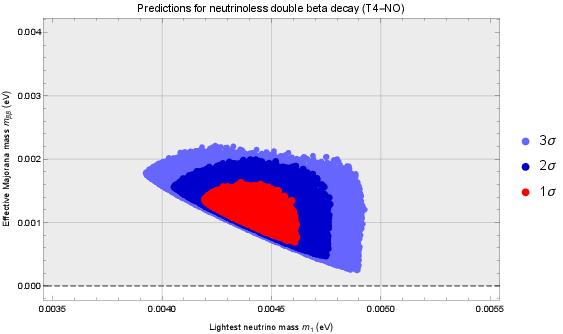}
        \caption{T4: $m_{\beta\beta}$ predictions.}
        \label{fig:T4_db}
    \end{subfigure}
    
    \caption{Comprehensive numerical results for the three independent neutrino mass textures (T2, T3, and T4) under Normal Ordering (NO). Left panels show the Dirac CP-violating phase $\delta_{\text{CP}}$ as a function of the lightest mass $m_1$. Middle panels display the correlation in the atmospheric parameter space ($\sin^2\theta_{23}$ vs. $\delta_{\text{CP}}$). Right panels illustrate the predictions for the effective Majorana mass $m_{\beta\beta}$ superimposed on the standard $3\sigma$ (NO) analytical band (gray area). Color-coded points represent $1\sigma$ (red), $2\sigma$ (dark blue), and $3\sigma$ (light blue) confidence levels.}
    \label{fig:global_panel_textures}
\end{figure*}

\subsection{T1 Inverted Ordering (IO)}

In the inverted ordering (IO) scenario, we find three texture-zero configurations capable of successfully reproducing the current neutrino oscillation data. In the following, we present the corresponding numerical results for the Dirac mass matrix \(M_D\), the Majorana mass matrix \(M_R\), the light-neutrino masses, and the neutrino oscillation parameters in the IO scheme, including the allowed Dirac CP-violating phase.

A viable benchmark is:

\begin{equation}
M_R =
\begin{pmatrix}
5.103\times10^{14} & 0 & 0 \\
0 & 8.254\times10^{14} & 0 \\
0 & 0 & 2.566\times10^{14}
\end{pmatrix} \text{ GeV},
\end{equation}

\begin{equation}
M_D =
\begin{pmatrix}
0 & 13.374+52.122i & -18.868-42.009i \\
13.374-52.122i & 0 & 12.809+10.519i \\
-18.868+42.009i & 12.809-10.519i & 0
\end{pmatrix} \text{ GeV}.
\end{equation}

\begin{align}
\sin^2\theta_{12} &\approx 0.275, \,\,\,\,
\sin^2\theta_{13} \approx 0.0241, \,\,\,\,
\sin^2\theta_{23} \approx 0.440, \\
\Delta m_{21}^2 &\approx 7.97\times10^{-5}\ \text{eV}^2, \,\,\,\,
\Delta m_{32}^2 \approx -2.55\times10^{-3}\ \text{eV}^2,\,\,\,\,
\delta_{CP} \approx 207^\circ.\nonumber
\end{align}

And the light-neutrino masses 

\begin{equation}
(m_1,m_2,m_3)\approx(0.0497,\ 0.0505,\ 0.0011)\ \text{eV}.
\end{equation}

Additionally, the results obtained for the three viable configurations are summarized in the following table \ref{Table2}. A more detailed numerical analysis, including the complete set of fitted parameters and mixing matrices, can be found in Appendix B.

\begin{table}[h]
\centering
\caption{Summary of viable inverted-ordering benchmark points.}
\begin{tabular}{c c c c c c c c}
\toprule
Texture & $\sin^2\theta_{12}$ & $\sin^2\theta_{13}$ & $\sin^2\theta_{23}$ & $\delta_{CP}$ & $\Delta m_{21}^2$ & $\Delta m_{32}^2$ & $\chi_{min}^2$ \\
\midrule
T1 & 0.304 & 0.0222 & 0.5751 & $60.52^\circ$ & $7.42\times10^{-5}$ & $2.49\times10^{-3}$ & $4.017\times10^{-6}$ \\
T2 & 0.304 & 0.0222 & 0.5750 & $89.16^\circ$ & $7.42\times10^{-5}$ & $2.49\times10^{-3}$ & $2.458\times10^{-9}$  \\
T4 & 0.304 & 0.0221 & 0.5236 & $263.55^\circ$ & $7.42\times10^{-5}$ & $2.49\times10^{-3}$ & $4.25$ \\
\bottomrule
\end{tabular}
\label{Table2}
\end{table}

All the obtained results are consistent within the \(3\sigma\) ranges reported by the NuFIT global analysis \cite{NuFIT}. Furthermore, although it was not imposed as an explicit fitting condition, the resulting neutrino mass spectra are also compatible with the current cosmological upper bound on the sum of the active neutrino masses \cite{Planck:2018vyg,DiValentino:2021hoh}.

Furthermore, a comparative statistical analysis between both mass orderings reveals a strong model preference towards the Normal Ordering (NO) scenario. Under identical Monte Carlo initialization spaces ($10^7$ sampled configurations), the total number of accepted points for the Inverted Ordering (IO) exhibits a reduction of approximately $10\%$ compared to NO, yielding only around $15,000$ viable parameter sets at the $3\sigma$ level. This significant decrease in sampling efficiency indicates that accommodating the quasi-degenerate and heavy mass spectrum typical of IO under the strict constraints of the proposed Dirac texture demands a substantially higher degree of parameter fine-tuning, thereby highlighting NO as the highly favored and more natural physical scenario within our theoretical framework.

\begin{figure*}[p] 
    \centering
    
    \begin{subfigure}[b]{0.32\textwidth}
        \centering
        \includegraphics[width=\textwidth]{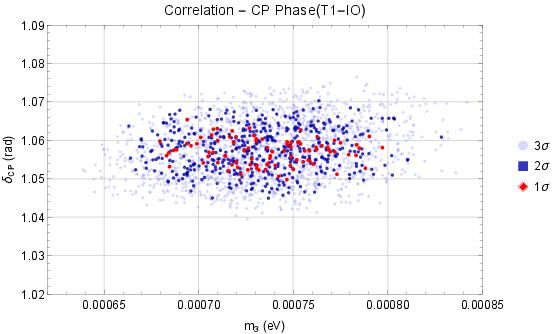}
        \caption{T1: $\delta_{\text{CP}}$ vs. $m_3$.}
        \label{fig:T1_IO_cp_m3}
    \end{subfigure}
    \hfill
    \begin{subfigure}[b]{0.32\textwidth}
        \centering
        \includegraphics[width=\textwidth]{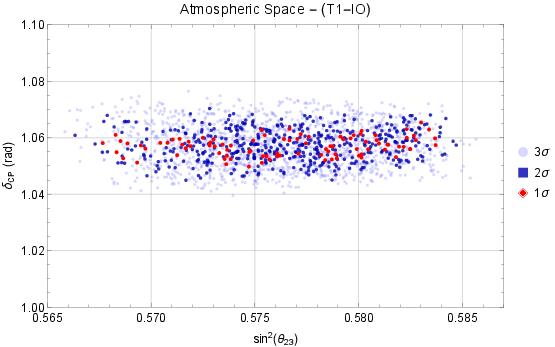}
        \caption{T1: $\sin^2\theta_{23}$ vs. $\delta_{\text{CP}}$.}
        \label{fig:T1_IO_cp_t23}
    \end{subfigure}
    \hfill
    \begin{subfigure}[b]{0.32\textwidth}
        \centering
        \includegraphics[width=\textwidth]{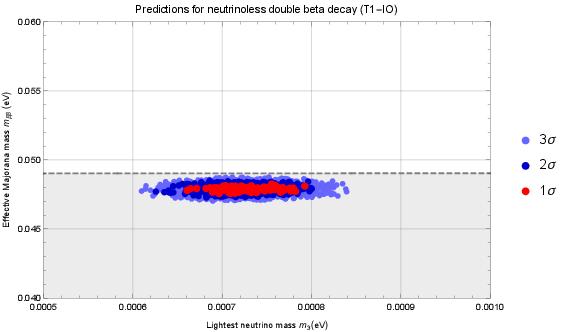}
        \caption{T1: $m_{\beta\beta}$ predictions.}
        \label{fig:T4_db}
    \end{subfigure}
    
    \vspace{0.3cm} 
    
    \begin{subfigure}[b]{0.32\textwidth}
        \centering
        \includegraphics[width=\textwidth]{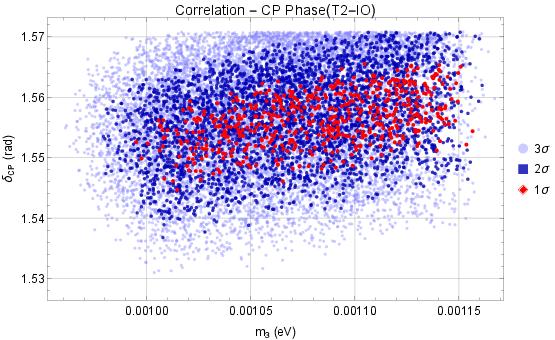}
        \caption{T2: $\delta_{\text{CP}}$ vs. $m_3$.}
        \label{fig:T2_IO_cp_m1}
    \end{subfigure}
    \hfill
    \begin{subfigure}[b]{0.32\textwidth}
        \centering
        \includegraphics[width=\textwidth]{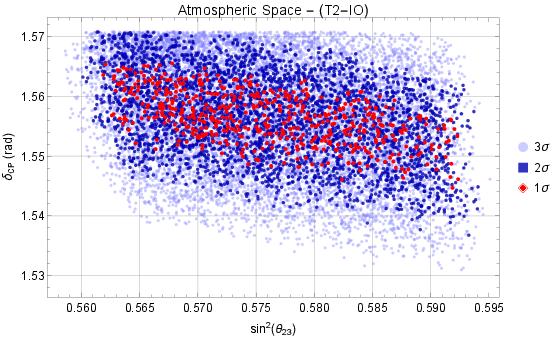}
        \caption{T2: $\sin^2\theta_{23}$ vs. $\delta_{\text{CP}}$.}
        \label{fig:T2_IO_cp_t23}
    \end{subfigure}
    \hfill
    \begin{subfigure}[b]{0.32\textwidth}
        \centering
        \includegraphics[width=\textwidth]{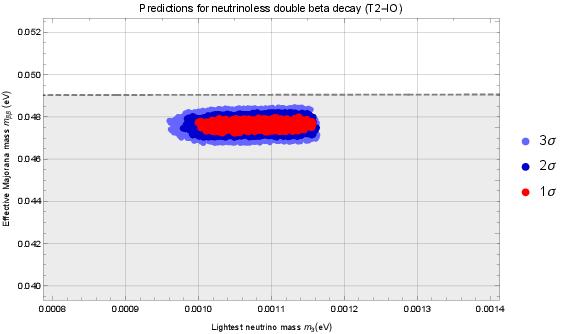}
        \caption{T2: $m_{\beta\beta}$ predictions.}
        \label{fig:T2_IO_db}
    \end{subfigure}
    
    \vspace{0.3cm} 
    
    \begin{subfigure}[b]{0.32\textwidth}
        \centering
        \includegraphics[width=\textwidth]{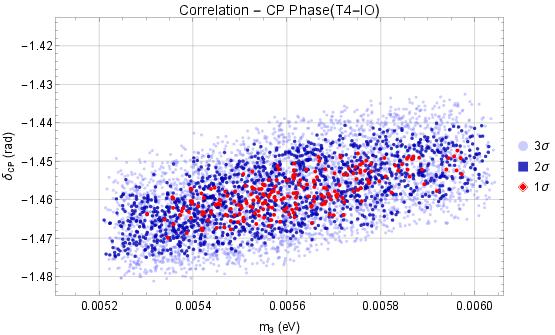}
        \caption{T4: $\delta_{\text{CP}}$ vs. $m_3$.}
        \label{fig:T4_IO_cp_m3}
    \end{subfigure}
    \hfill
    \begin{subfigure}[b]{0.32\textwidth}
        \centering
        \includegraphics[width=\textwidth]{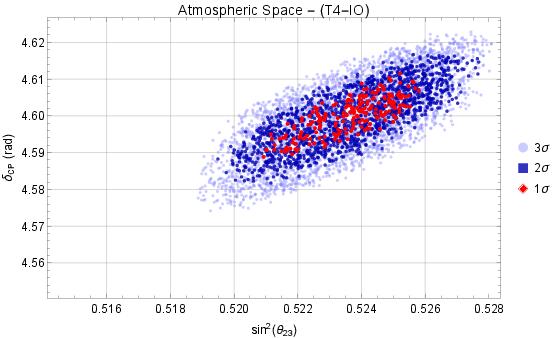}
        \caption{T4: $\sin^2\theta_{23}$ vs. $\delta_{\text{CP}}$.}
        \label{fig:T4_IO_cp_t23}
    \end{subfigure}
    \hfill
    \begin{subfigure}[b]{0.32\textwidth}
        \centering
        \includegraphics[width=\textwidth]{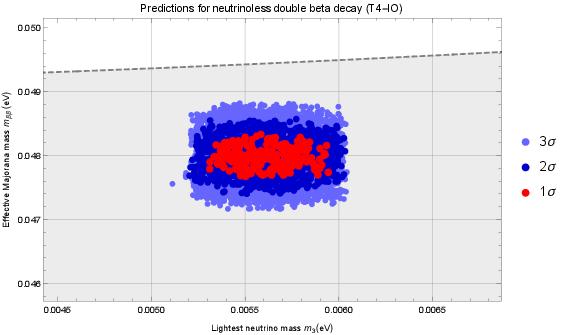}
        \caption{T4: $m_{\beta\beta}$ predictions.}
        \label{fig:T4_IO_db}
    \end{subfigure}
    
    \caption{Comprehensive numerical results for the three independent neutrino mass textures (T1, T3, and T4) under Inverted Ordering (IO). Left panels show the Dirac CP-violating phase $\delta_{\text{CP}}$ as a function of the lightest mass $m_1$. Middle panels display the correlation in the atmospheric parameter space ($\sin^2\theta_{23}$ vs. $\delta_{\text{CP}}$). Right panels illustrate the predictions for the effective Majorana mass $m_{\beta\beta}$ superimposed on the standard $3\sigma$ (IO) analytical band (gray area). Color-coded points represent $1\sigma$ (red), $2\sigma$ (dark blue), and $3\sigma$ (light blue) confidence levels.}
    \label{fig:global_panel_textures}
\end{figure*}

\section{Conclusions}

In this work, we have shown that it is possible to successfully reproduce all current physical observables related to neutrino physics by introducing texture zeros in the neutrino mass matrices within the framework of the Type-I seesaw mechanism. Our analysis is based on the assumption that an underlying model exists capable of generating light-neutrino masses through the seesaw mechanism, where the Dirac neutrino mass matrix is Hermitian, the Majorana mass matrix is diagonal, and the upper-left block of the complete neutrino mass matrix vanishes.

Previous studies have shown that, in the absence of Majorana neutrinos, that is, when neutrino masses are generated purely through a Dirac-type mass matrix, Hermitian mass matrices can accommodate at most two texture zeros while remaining compatible with neutrino oscillation data. In such scenarios, configurations containing three texture zeros fail to reproduce the observed physical parameters.

However, within the framework considered in this work, we have demonstrated that the implementation of the Type-I seesaw mechanism significantly modifies this conclusion. In particular, among the twenty possible Hermitian Dirac mass matrices containing three texture zeros, four viable configurations are found in the normal ordering (NO) scenario, while three viable textures are obtained for inverted ordering (IO). This becomes possible due to the structure imposed by the seesaw mechanism, where the Majorana mass matrix \(M_R\) corresponds to a high-energy scale much larger than the entries of the Dirac mass matrix \(M_D\), while the upper-left diagonal block remains null.

It is worth emphasizing that the textures denoted as \(T_1\), \(T_2\), and \(T_4\) are simultaneously viable in both normal and inverted ordering schemes. Consequently, the introduction of texture zeros considerably reduces the number of free parameters in the neutrino sector while still allowing the coexistence of both Dirac- and Majorana-type neutrino mass terms.

From the phenomenological point of view, the viable textures obtained in this analysis successfully reproduce the three leptonic mixing angles, the two neutrino mass-squared differences, and therefore the physical neutrino masses themselves. Furthermore, the framework allows the calculation of the Jarlskog invariant and, equivalently, the determination of the leptonic Dirac CP-violating phase. An additional important result is that all viable configurations naturally satisfy the current cosmological bounds on the sum of the active neutrino masses.

Overall, the present analysis demonstrates that texture-zero structures within the Type-I seesaw mechanism constitute a predictive and phenomenologically viable framework for describing neutrino masses and mixing, while substantially reducing the arbitrariness associated with the neutrino mass sector.

In general, $M_D$ in (1) is an arbitrary $3 \times 3$ mass matrix, with 18 free parameters and no predictive power at all. By assuming it is Hermitian, as in our analysis above, we make it predictable and with phenomenological implications. The analysis with a general $M_D$ is presented in Ref. \cite{Adhikary:2012zx}, where they use a $\mu$ - $\tau$ symmetry to reach phenomenological implications.

\section{Acknowledgments}
RB and JG wish to thank the ITM for providing the time necessary to carry out this project.

\section{Appendix}

\subsection{Appendix A}

\subsubsection{Texture T2: zeros in $(1,1)$, $(2,2)$ and $(2,3)$}

\begin{equation}
M_D^{\mathrm{T2}}=
\begin{pmatrix}
0 & 48.8232+27.2051i & 17.7813-59.1214i\\
 48.8232-27.2051i & 0 & 0\\
17.7813+59.1214i & 0 & 109.4900
\end{pmatrix}\ \mathrm{GeV},
\end{equation}

\begin{equation}
M_R^{\mathrm{T2}}=\mathrm{diag}\left(1.238\times10^{14},\ 2.586\times10^{14},\ 7.099\times10^{14}\right)\ \mathrm{GeV}.
\end{equation}

The light-neutrino masses are
\begin{equation}
(m_1,m_2,m_3)=(0.008410,\ 0.012053,\ 0.050778)\ \mathrm{eV},
\end{equation}
with
\begin{equation}
\sum_i m_i = 0.071242\ \mathrm{eV}.
\end{equation}

The oscillation observables are
\begin{align}
\sin^2\theta_{12} &= 0.303,\,\,\,\,
\sin^2\theta_{13} = 0.0219,\,\,\,\,
\sin^2\theta_{23} = 0.5374,\\
\Delta m_{21}^2 &= 7.45\times10^{-5}\ \mathrm{eV}^2,\,\,\,\,
\Delta m_{31}^2 = 2.51\times10^{-3}\ \mathrm{eV}^2,\,\,\,\,
\delta_{CP} = 264.88^\circ.\nonumber
\end{align}

\subsubsection{Texture T3: zeros in $(1,1)$, $(3,3)$ and $(1,2)$}

\begin{equation}
M_D^{\mathrm{T3}}=
\begin{pmatrix}
0 & 0 & 37.3401-56.3909i\\
0 & -67.5304 & -102.462-91.1035i\\
37.3401+56.3909i& -102.462+91.1035i & 0
\end{pmatrix}\ \mathrm{GeV},
\end{equation}

\begin{equation}
M_R^{\mathrm{T3}}=\mathrm{diag}\left(1.033\times10^{14},\ 3.341\times10^{14},\ 1.000\times10^{15}\right)\ \mathrm{GeV}.
\end{equation}

The light-neutrino masses are
\begin{equation}
(m_1,m_2,m_3)=(0.005397,\ 0.010165,\ 0.050390)\ \mathrm{eV},
\end{equation}
with
\begin{equation}
\sum_i m_i = 0.065953\ \mathrm{eV}.
\end{equation}

The oscillation observables are
\begin{align}
\sin^2\theta_{12} &= 0.304,\,\,\,\,
\sin^2\theta_{13} = 0.0222,\,\,\,\,
\sin^2\theta_{23} = 0.5700,\\
\Delta m_{21}^2 &= 7.42\times10^{-5}\ \mathrm{eV}^2,\,\,\,\,
\Delta m_{31}^2 = 2.51\times10^{-3}\ \mathrm{eV}^2,\,\,\,\,
\delta_{CP} = 231.42^\circ.\nonumber
\end{align}

\subsubsection{Texture T4: zeros in $(1,1)$, $(3,3)$ and $(2,3)$}

\begin{equation}
M_D^{\mathrm{T4}}=
\begin{pmatrix}
0 & 43.1285+i & 1-49.8243i\\
43.1285-i & 46.1151 & 0\\
 1+49.8243i & 0 & 0
\end{pmatrix}\ \mathrm{GeV},
\end{equation}

\begin{equation}
M_R^{\mathrm{T4}}=\mathrm{diag}\left(9.861\times10^{13},\ 2.031\times10^{14},\ 3.054\times10^{14}\right)\ \mathrm{GeV}.
\end{equation}

The light-neutrino masses are
\begin{equation}
(m_1,m_2,m_3)=(0.004397,\ 0.009705,\ 0.050249)\ \mathrm{eV},
\end{equation}
with
\begin{equation}
\sum_i m_i = 0.0643512\ \mathrm{eV}.
\end{equation}

The oscillation observables are
\begin{align}
\sin^2\theta_{12} &= 0.306,\,\,\,\,
\sin^2\theta_{13} = 0.0221,\,\,\,\,
\sin^2\theta_{23} = 0.5692,\\
\Delta m_{21}^2 &= 7.49\times10^{-5}\ \mathrm{eV}^2,\,\,\,\,
\Delta m_{31}^2 = 2.51\times10^{-3}\ \mathrm{eV}^2,\,\,\,\,
\delta_{CP} = 182^\circ.\nonumber
\end{align}

\subsection{Appendix B}

\subsubsection{T2 in Inverted Ordering}

\[
M_D=
\begin{pmatrix}
0 &61.676+82.414i & 27.991+120.446i\\
61.676-82.414i & 0 & 0\\
27.991-120.446i & 0 & -53.779
\end{pmatrix}
\]

\[
M_R=
\mathrm{diag}(5.037\times10^{14},\,3.912\times10^{14},\,6.135\times10^{14})
\]

The light-neutrino masses are

\begin{align}
(m_1, m_2, m_3 )&= (0.04916, 0.04991,0.00109)\ \text{eV}, \\
\sum m_i &= 0.10017\ \text{eV}.
\end{align}

\begin{align}
\sin^2\theta_{12} &= 0.304, \,\,\,\,
\sin^2\theta_{13} = 0.02225, \,\,\,\,
\sin^2\theta_{23} = 0.575,\\
\Delta m_{21}^2 &= 7.42\times10^{-5}\ \text{eV}^2, \,\,\,\,
\Delta m_{32}^2 = -2.49\times10^{-3}\ \text{eV}^2,\,\,\,\,\delta_{CP} = 90.83^\circ\nonumber
\end{align}

\subsubsection{T4 in Inverted Ordering}

\[
M_D=
\begin{pmatrix}
0 & 78.248-64.803i & 106.568+54.1011i\\
78.248+64.803i & 85.930 & 0\\
 106.568-54.1011i & 0 & 0
\end{pmatrix}
\]

\[
M_R=
\mathrm{diag}(5.103\times10^{14},\,8.254\times10^{14},\,2.566\times10^{14})
\]

\begin{align}
(m_1, m_2,m_3) &= (0.04946,0.05021,0.00561)\ \text{eV}, \\
\sum m_i &= 0.105294\ \text{eV}.
\end{align}

\begin{align}
\sin^2\theta_{12} &= 0.304, \,\,\,\,
\sin^2\theta_{13} = 0.02216, \,\,\,\,
\sin^2\theta_{23} = 0.52359,\\
\Delta m_{21}^2 &= 7.422\times10^{-5}\ \text{eV}^2, \,\,\,\,
\Delta m_{32}^2 = 2.49\times10^{-3}\ \text{eV}^2,\,\,\,\,
\delta_{CP} = 263.547^\circ.\nonumber
\end{align}

\bibliographystyle{unsrt}
\bibliography{referencias} 

\end{document}